\documentstyle{lamuphys}
\makeatletter
\let\chapter\hid@chapter
\makeatother
\begin{document}
\pagenumbering{arabic}
\title{Susy Hierarchies and Affine Lie Algebras}

\author{Francesco\,Toppan}

\institute{Shizuoka University, Department of Physics,
Ohya 836, Shizuoka city, Japan}

\maketitle

\begin{abstract}
We review some basic features of the Lie-algebraic classification
of $W$-algebras and related integrable hierarchies in $1+1$ dimensions,
pointing out the
role of affine Lie algebras. We emphasize that the supersymmetric
extensions of the above construction possibly lead, though some questions
are still opened, to the classification of supersymmetric hierarchies
based on ``generic'' supersymmetric affine Lie algebras. Here the word
generic is used to make clear that well-known procedures, as those
introduced by Inami and Kanno, are too restricted and do not lead to
the full spectrum of supersymmetric integrable hierarchies one can construct.
A particular attention is devoted to the large-$N$ supersymmetric extensions
(here $N=4$). The attention paid by large-$N$ theories being due to the
fact that they arise as dimensional reduction of $N=1$ models, and moreover
that they realize an ``unification'' of known hierarchies.
\end{abstract}
\section{Introduction}
In the last few years a lot of attention has been devoted to inter-related
topics which go under the name of $W$-algebras, integrable hierarchies
(non-relativistic) in $1+1$ dimensions of KdV or NLS type, $2$-dimensional
reativistic equations like Liouville (more generally Toda field theories)
or SG, $0$-dimensional matrix models which describe discretized $2$-dimensional
gravity.\par
Some of the above topics have definitely a more mathematical flavour, like for
instance 
the theory and classification of $W$-algebras; some others are definitely
more physically grounded, KdV describes waves in shallow water, Liouville
equation is an ubiquous one, but at least a lot of attention has been paid
to it in connection with non-critical strings.\par
Despite the fact that the above models and theories seem all very different
and can be constructed in apparently unrelated ways it turned out indeed
that they are just manifestations of an underlying mathematical framework.
Indeed $W$-algebras (i.e. non-linear Poisson-bracket algebras of
$1$-dimensional fields
containing a Virasoro one, which
satisfy the standard properties of antisymmetry, Jacobi identity and
Leibniz rule in the classical case) turn out to be the Poisson bracket
structures for both relativistic Toda field theories and non-relativistic
integrable equations like KdV, Boussinesq and so on. Moreover the Ward
identities of generalized matrix models generate the so-called W-constraints
and their partition functions turn out to be related to the
$\tau$-functions of associated classical integrable hierarchies\cite{df}.\par
It deserves being mentioned that $W$-algebras themselves can be produced
and classified via a truly algebraic approach, by putting restrictions to
affine Lie algebras; such restrictions can be realized either as hamiltonian
reductions or coset constructions (by looking at some centralizer over some
enveloping algebra \cite{d}).
While there is maybe no strict mathematical proof so far
that all $W$-algebras can be obtained with the methods of \cite{fss},
at least there is
no need to believe that all $W$-algebras cannot be obtained that way.\par
The production of such a closed structure like a $W$-algebra is an interesting
mathematical activity by itself, however there is much more than that. A very
peculiar and absolutely non-trivial feature of $W$-algebras arises
when they allow constructing towers of infinite hamiltonians in involution.
In this way they turn out to be linked to a dynamical system of a special
kind, an integrable one. The technical tool which allows to prove integrability
consists in
formulating the dynamical system (and its associated $W$-algebra) via a Lax
operator which can be either of scalar (KP-like) or matrix type.\par
In
the next section we will sketch the main features of the bosonic construction,
postponing to the later section the introduction of supersymmetric integrable
systems with the necessary modifications.

\section{Bosonic Hierarchies}
Let us first point out that $2D$ relativistic Toda models and $1+1$
non-relativistic integrable equations arise from constraining affine Lie
algebras ${\hat{\cal G}}$ (and their associated enveloping algebras).
The basic difference in the relativistic case is
due to the fact that two copies of the affine algebra are considered,
associated to the chiral and antichiral currents
$J(z)$, ${\overline J}({\overline z})$ respectively. The dynamical fields
are group-valued $g(z,{\overline z})$ and possibly expressed through a
Gauss decomposition. We have
\begin{eqnarray}
J(z)= g^{-1}\partial_z g
\end{eqnarray}
and a similar equation for ${\overline J}({\overline z})$.\par
The simplest case is provided when ${\cal G}=sl(2)$. The three currents
associated
to ${\hat {sl(2)}}$ are $J_{\pm}(x)$ and $J_0(x)$ ($J_0(x)$ generates the
${\hat{U(1)}}$ subalgebra).\par
In this simple case only two inequivalent constraints can be imposed on the
(enveloping) affine algebra, either\par
A) constraining $J_+(x) =1$ (hamiltonian constrain), or\par
B) selecting the $X(y)$ centralizer of the enveloping algebra, namely\par
$\{J_0(x), X(y)\}=0$ (coset).\par
Accordingly, we get in the relativistic (I) and in the non-relativistic
(II)
cases
the following dynamical systems:\par
I A) The Liouville equation.\par
II A) The m-KdV (and KdV) equation.\par
I B) The $2D$ Witten`s black hole.\par
II B) The Non-Linear Schr\"odinger Equation.\par
From now on we will concentrate only on the non-relativistic case, that is
the system of integrable equations in $1+1$ dimension which can be solved
through
inverse scattering method. As mentioned above the integrability property
is expressed by the fact that one can express the equations of motion through
a Lax operator. We have two kinds of such operators, the scalar type
\begin{eqnarray}
L=\partial +\sum_{i=1,...,\infty} u_i \partial^{-i}
\end{eqnarray}
associated to the KP hierarchy, and the matrix type
\begin{eqnarray}
{\cal L} = \partial + \sum_i J_i(x) \tau^{i} + \Lambda
\end{eqnarray}
where the currents $J_i(x)$ are valued in some Lie algebra ${\cal G}$ generated
by $\tau^i$. $\Lambda$ is a constant element, depending on a spectral parameter
$\lambda$, such that the loop algebra ${\tilde G} = {\cal G}\otimes C(\lambda,
\lambda^{-1})$ can be decomposed in the direct sum
${\tilde G} = K \oplus M$,
with $K$, $M$ respectively the Kernel and Image under
the adjoint action
of $\Lambda$ (this technical property implies that ${\cal L}$
can be diagonalized under a similarity transformation).\par
In order to extract from scalar Lax operators integrable equations involving
only
a finite number of fields, we have to constrain the infinite fields $u_i(x)$
in a way consistent with the KP flows (constrained KP hierarchies). One
possibility is requiring e.g. for a given $n$  $L^n={L^n}_+$ (that is to be a
purely differential operator). This
is indeed a consistent constraint (leading to
the n-th KdV hierarchies), however it is known there exists many more
inequivalent consistent constraints and a classification of them out of the
scalar Lax operators alone appears rather impractical.\par
On the contrary it is well-known how to classify all possible hierarchies
associated with affine algebras. They turn out to be related to the acceptable
integral grading for any given loop algebra ${\cal G}$ and the choice of
the regular element
$\Lambda$ (see e.g. \cite{mat} for details).
Moreover it is possible to relate such
solutions with the constrained KP hierarchies. \cite{chi}
\par
In
order to be explicit we recall that in the original Drinfeld-Sokolov paper the
n-th KdV hierarchies were obtained by assuming the underlying algebra to be
$sl(n)$ and the regular element $\Lambda$ to be the
sum over the ${\tilde{sl(n)}}$
simple roots.\par
The above scheme seems quite satisfactory from the point of view of bosonic
hierarchies since it provides a well-defined construction for them and is quite
plausible they can all be accomodated in it. Questions concerning the
possible equivalence of hierarchies arising from different choices of algebras,
integral grading and/or regular
element seem more tehnical and less central.\par
So far for purely bosonic hierarchies, in the next section we will introduce
the supersymmetric ones.

\section{Supersymmetric Hierarchies}
The
first natural question when discussing supersymmetric integrable hierarchies
is of course why should we worry about them.
One can think e.g. to the fact that so far no
supersymmetric matrix model providing a discretized $2D$ supergravity has been
produced. Neverthless some achievement has been made like the introduction of
a supereigenvalue model which is in a sense pull out of a hat but is related to
a superintegrable hierarchy\cite{sem}.
\par
Morever the remarkable relation of
KdV-type hierarchies with the conformal algebras
(Virasoro and supersymmetric extensions) establishes a connection between such
hierarchies and the (super-)string theories which has still to be fully
appreciated.\par
From a purely mathematical point of view the role of supersymmetric integrable
hierarchies and superalgebras is essential in at least two respects.
Indeed, even when considering purely bosonic hierarchies, if we not allow
for super-structures like super-algebras we cannot pretend to exhaust the
full class of possible hierarchies, new integrable interacting purely bosonic
hierarchies arise in fact from the bosonic sector (B-B and F-F submatrices)
of supermatrix-valued superhierarchies.\par
Moreover investigating large $N$-extended superhierarchies corresponds to a
sort of ``unification or grandunification'' program of known hierarchies.
It happens in fact that unrelated bosonic hierarchies or lower supersymmetric
($N=1,2)$ hierarchies turn out to be different manifestations of a single
``unifying''
large $N$ supersymmetric hierarchy. We will see later an example of
this fact when discussing the $N=4$ KdV hierarchy.\par
The point of view that we adopt here in discussing supersymmetric
hierarchies is based on the (super-)Lie algebra framework, which one can
reasonably hope will provide the key to classify all superhierarchies.
The
main reason of the difficulty involved in classifying superhierarchies w.r.t.
the purely bosonic ones is due to the complications involving the presence of
both even and odd generators.\par
We need to point out that (super)-Lie algebras appear in $3$ different classes
according whether they admit a presentation in terms of Dynkin diagrams
with simple roots which are either:\par
i) purely fermionic,\par
ii) necessarily of mixed type, or\par
iii) purely bosonic (they are reduced to standard Lie algebras).\par
A simple argument made people believe for a long time that only the special
class of super-Lie algebras admitting fermionic simple roots were relevant for
the construction of superhierarchies. Inami and Kanno \cite{ik}
gave it in the contest
of super-KdV hierarchies. In order to extend the bosonic matrix Lax operator
they were led to consider a supersymmetric Lax of the kind
\begin{eqnarray}
{\cal L} = D + \Psi(X) + \Lambda
\end{eqnarray}
where now $D$ is the ($N=1$) fermionic supersymmetric derivative.
$\Psi(X)$ are
superfields valued in some superalgebra and as such are fermionic.
The regular element $\Lambda$ should be given by the sum over the simple roots
and in order to respect statistics it must be fermionic as well.
Therefore it seemed that only class i) superalgebras had to be considered.
A similar argument was given by Evans and Hollowood \cite{eh}
in the case of superToda
theories. This argument has paved the way to the standard supersymmetrization
recipe of bosonic models which goes as follows:
embed the given bosonic algebra which provides the bosonic system into a larger
superLie algebra
having ``good properties'' and perform the hamiltonian reduction
on it. In this way $N=1$ extension of KdV and Liouville equations were provided
in terms of the $osp(1|2)$ superalgebra.\par
The Inami-Kanno scheme
is a perfectly consistent one, leading to a classification
of supersymmetric hierarchies much similar to the bosonic case. There would be
no need to look for improving it if it would not turn out a too restricted one.
Indeed
it happens
that well-known and interesting superhierarchies cannot be accomodated
in it. To my knowledge Brunelli and Das were the first \cite{bd}
who faced this problem
when they realized that the superNLS equation admits a Lax operator based on
the $sl(2)$ bosonic algebra (and not $osp(1|2)$ as one would have been
expected). In \cite{sns}
it was pointed out that the superNLS hierarchy arises as a
coset construction (just as its bosonic counterpart) over a Poisson bracket
structure based on the superaffinization (that is expressed in terms of
superfields) of the bosonic $sl(2)$ algebra.
\par
As a consequence there exists a much bigger class of supersymmetric
integrable models than previously expected which need to be investigated.
Despite the fact that we do not have yet a systematic way of constructing them
in terms
of matrix Lax operators just like the bosonic models or the Inami-Kanno
superhierarchies, still we can develop some strategy to investigate them.
This will be explained next.

\section{Supersymmetric Hierarchies and Affine Algebras}
Let us here discuss a possible strategy for constructing supersymmetric
integrable hierarchies from
generic superaffinizations of (super-)Lie algebras.  But first let us point out
that a superaffinization of a given (super-)Lie algebra
${\cal G}$ with generators $\tau^i$ and structure constants ${f^{ij}}_k$
is realized by $N=1$ superfields $\Psi^i(X)$, with opposite statistics w.r.t.
$\tau^i$ and such that
\begin{eqnarray}
\{\Psi^i(X),\Psi^j(Y)\}=
{f^{ij}}_k\Psi^k(Y)\delta(X,Y) + K^{ij} D_Y\delta (X,Y)
\end{eqnarray}
with $\delta(X,Y)$ the $N=1$ delta-function and $K^{ij}=Str({\tau^i\tau^j})$
in some given (let's say the adjoint) representation of ${\cal G}$.\par
The following steps should be performed:\par
i) Take a superaffine (super-)Lie algebra which should be regarded as Poisson
bracket structure.\par
ii) Make some Ansatz over the possible hamiltonians in involution; this would
mean imposing symmetry requirements, cosets or hamiltonian reductions.\par
iii) Check the consistency of flows and if indeed at lower orders one gets
hamiltonians in involution.\par
iv) Try to figure out the form of possible Lax operators (this is the most
difficult task).\par
It can even happen that one finds more structures than expected. Indeed it is
well-known that a relation exists between division algebras ad extended
supersymmetries.
Complex, Quaternionic ad Octonionic structures are associated to (global)
$N=2,4,8$ extensions respectively.\par
A complex structure for a (super-)algebra over the real fields is an operation
$J$ which satisfy $J^2=-1$, while a quaternionic structure involves $3$ complex
structures $J_i$ whose mutual algebra is that of the Pauli matrices.\par
If a theory admits a complex structure it necessarily has an extended
supersymmetry. For instance the superNLS equation which arises from the
${\hat {sl(2)}}/ {\hat{u(1)}}$
structure is automatically $N=2$ since such a coset
admits a complex structure (while ${\hat{sl(2)}}$ does not). An elegant
(but equivalent) formulation can be realized through the coset
${\hat {sl(2)\oplus u(1)}}/{\hat{u(1)\oplus u(1)}}$ \cite{kst}.
This construction allows a
manifestly $N=2$ superfield formulation since the extra $u(1)$ are added to
give a complex structures for both numerator and denominator.\par
The $sl(2)\oplus u(1)$
algebra turns out to be vey interesting because it appears
in the list given by \cite{xxx}
(actually these authors considered group-manifolds, out
of which algebras can be immediately recovered) as the simplest example of
non-abelian algebra (the even simpler abelian case being $u(1)^{\otimes 4}$)
admitting a quaternionic structure.\par
A natural question therefore arises, namely if it is
possible that the
superaffine algebra ${\hat{sl(2)\oplus u(1)}}$, taken as a Poisson bracket
algebra, would
allow to play another game, not just the coset already mentioned,
according to the above scheme. In particular we can ask ourselves
if we can demand an $N=4$ symmetry requirement
which in turns imply an $N=4$ hierarchy. In the next section we
will show that this is indeed the case \cite{ikt}.

\section{the $N=4$ structure of ${{sl(2)\oplus u(1)}}$}

The superaffine algebra ${\hat{sl(2)\oplus u(1)}}$ can be conveniently
described in terms of $N=2$ superfields. Let us introduce the $N=2$
fermionic derivatives $D,{\overline D}$ whose algebra reads as follows
\begin{eqnarray}
D^2={\overline D}^2 &=&0 \nonumber\\
\{D,{\overline D}\}&=& -\partial_x
\end{eqnarray}
The spin $\frac{1}{2}$ $N=2$ superfields are denoted as $H, {\overline H},
F, {\overline F}$. \par
$H$ and ${\overline H}$ are associated to the $u(1)\oplus
u(1)$ subalgebra.\par
They are constrained superfields, the constraints being non-linearly realized
\begin{eqnarray}
&& DH = {\overline D}{\overline H}=0\nonumber\\
&& (D+ H) F = ({\overline D} -{\overline H}){\overline F} =0
\end{eqnarray}
The non-vanishing structure constants are given by
\begin{eqnarray}
&&\{ H (1),{\overline H}(2)\} = D{\overline D} \delta \nonumber\\
&&\{ H(1), F(2)\} = DF\cdot \delta\nonumber\\
&& \{H(1), {\overline F}(2)\} = - D{\overline F} \delta\nonumber\\
&&\{{\overline H}(1), F(2)\} = -{\overline D} F\cdot \delta\nonumber\\
&&\{{\overline H}(1), {\overline F}(2)\} = {\overline D} {\overline F} \cdot
\delta\nonumber\\
&&\{ F(1),{\overline F}(2)\} = (D+H) ({\overline D} +{\overline H}) \delta
+ F {\overline F}\delta
\end{eqnarray}
where $\delta\equiv \delta (1,2)$
is the $N=2$ delta-function and the derivatives
in the r.h.s. are computed at $Z\equiv 1$.
\par
In the last line a ``fake'' non-linear term appears. It is not present when
the chiral constraints are
solved in terms of $N=1$ superfields or component fields.\par
There exists a second set of {\it global} $N=2$ non-linear supersymmetries,
expressed through the infinitesimal parameters $\epsilon, {\overline\epsilon}$,
which results from the quaternionic structure
associated with $sl(2)\oplus u(1)$.
We have
\begin{eqnarray}
\delta H &=& \epsilon D{\overline F} + {\overline \epsilon} H F\nonumber\\
\delta{\overline H} &=& {\overline \epsilon} {\overline D}F - \epsilon
{\overline H}{\overline F}\nonumber\\
\delta F &=& -\epsilon D
{\overline H} -\epsilon (H{\overline H} + F{\overline F})
\nonumber\\
\delta{\overline F} &=& -{\overline \epsilon}
{\overline D} H - {\overline \epsilon}
(H{\overline H} + F{\overline F})
\end{eqnarray}
It can be easily checked that the above transformations preserve the
chirality constraints and that their commutators close to give,
together with the original transformations, an $N=4$ supersymmetry.

\section{the $N=4$ Hierarchy}

We have seen that ${\hat{sl(2)\oplus u(1)}}$ carries an $N=$4 structure.
To prove the existence of globally invariant $N=4$ dynamical systems we have
to construct explicitly the $N=4$ invariant hamiltonians.
They indeed exist and moreover, at the lower
dimensional integral spin dimension
$d=1,2$, they are unique up to total derivatives
(at least if a global chargeless
condition is required, where $H, {\overline H}$ are chargeless, while $F$ and
${\overline F}$ have charges $+1$ and $-1$ respectively).\par
We have indeed
\begin{eqnarray}
H_1 &=& F {\overline F} + H {\overline H}\nonumber\\
H_2 &=& F'{\overline F} -H'{\overline H} -\nonumber\\
&& -(D{\overline H} + {\overline D} H) (H {\overline H} + F {\overline F})
-2 H {\overline H} F{\overline F}
\end{eqnarray}
Higher dimensional $N=4$ hamiltonians can be explicitly constructed and turn
out to be in involution with the lower dimensional ones.\par
The resulting equations of motion (which is not needed to write here,see
\cite{ikt})
with respect to the
second hamiltonian realize an $N=4$ dynamical system which combines in a
non-trivial way both the $N=2$ mKdV equation and the $N=2$ NLS equations.
The latters are recovered by setting, consistently with the equations of
motion,
respectively $F={\overline F} =0$ and $H={\overline H} =0$. A third,
more mysterious, $N=2$ system can be obtained by performing a non-symmetrical
reduction leading to ${\overline H} = F = 0$.\par
So far for what concerns the construction of the $N=4$ system. An Ansatz has
guided us towards its realization and we have seen that it is essentially
unique.
A point which has been left apart consists in explicitly proving that our
system
indeed corresponds to an integrable hierarchy admitting an infinite tower of
hamiltonians in involution. In this particular case
we have a very elegant procedure
which proves that. Unfortunately, as already stated, we cannot rely
so far on any systematic construction for the Lax operators valid for generic
theories. The best we can do at present is based on a trial procedure.\par
However, for the ${\hat{sl(2)\oplus u(1)}}$ case the key property which allows
to solve the problem is the existence of a (differential polynomial) $N=4$
Sugawara construction \cite{ikt}.
This very remarkable transformation has at least four
different consequences:\par
i) it provides a linearization of the $N=4$ transformations,\par
ii) it furnishes a realization for the ``minimal'' $N=4$ SuperConformal Algebra
(SCA),\par
iii) it relates the ``affine hierarchy'' to the $N=4$ KdV system\cite{di}
and\par
iv) it allows the construction of the Lax operator which proves the
integrability.
\par
The Sugawara transformation is a differential polynomial transformation which
express the ``superconformal fields'' through the original affine
superfields $H, {\overline H}, F, {\overline F}$. The transformed superfields
are an $N=2$ real superVirasoro superfield $J$ (with component fields content
$(1,\frac{3}{2},\frac{3}{2},2)$) plus two chiral and antichiral spin $1$
superfields (in components $(1,\frac{3}{2})$). We have explicitly
\begin{eqnarray}
J &=& H{\overline H} + F{\overline F} + D {\overline H} + {\overline D} H
\nonumber\\
\Phi &=& D {\overline F}\nonumber\\
{\overline \Phi} &=& {\overline D}F
\end{eqnarray}
The presence of the Feigin-Fuchs terms in the r.h.s. for $J$ is especially
important. Without them $J$ would be a nilpotent field ($J^3=0$ due to the
fermionic character of $H, {\overline H}, F, {\overline F}$). Moreover they
allow the second set of $N=2$ transformations to close linearly on $J, \Phi,
{\overline \Phi}$ as
\begin{eqnarray}
\delta J &=& -{\overline \epsilon} D{\overline \Phi} -{\epsilon}{\overline D}
\Phi\nonumber\\
\delta \Phi &=& {\overline \epsilon} D J\nonumber\\
\delta {\overline \Phi} &=& \epsilon {\overline D}\Phi
\end{eqnarray}
The composite superfields $J, \Phi, {\overline \Phi}$ satisfy a closed algebra
structure under the original ${\hat{sl(2)\oplus u(1)}}$ Poisson bracket
structure
and it coincides with the minimal version of the $N=4$ SCA.
\par
The hamiltonians in involution can be closely expressed through the superfields
$J, \Phi,{\overline \Phi}$ alone. At the lowest order we have, for the
hamiltonian
densities
\begin{eqnarray}
H_1 &=& J\nonumber\\
H_2&=& J^2 - 2 \Phi{\overline\Phi}\nonumber\\
H_3&=& J[D,{\overline D}] J + 2 \Phi {\overline \Phi}'+\frac{2}{3}J^3
- 4 J \Phi{\overline \Phi}
\end{eqnarray}
As a consequence we have a closed system of dynamical equations for
$J,\Phi ,{\overline\Phi}$ which coincides with the $N=4$ KdV hierarchy.
\par
The Lax operator can be borrowed from the known Lax operator of KdV and is
given
by \cite{dig}
\begin{eqnarray}
L= D {\overline D} + D {\overline D} \partial^{-1}(J +{\overline \Phi}
\partial^{-1}\Phi ) \partial^{-1} D {\overline D}
\end{eqnarray}
It should be noticed that in this particular case checking the integrability
properties of the given hierarchy was immediate once the Sugawara construction
has been taken into account since the above Lax operator for the $N=4$ KdV
was already known. However, even if this would have not been the  case
(as one could expect
from constructions based on more general algebras), the Sugawara transformation
itself would greatly simplify the task of finding the correct Lax pair, since
it is much easier to deal with three spin $1$ fields than with spin
$\frac{1}{2}$ superfields. The dramatic simpification of the hamiltonians
when expressed through $J, \Phi, {\overline\Phi}$ is also an example.\par
Some more comments are in order: the $N=4$ KdV is the ``unifying hierarchy''
for two of the three inequivalent $N=2$ KdV hierarchies labeled by
$a= 1,-2,4$. The $a=-2$ and $a= 4$ N=2 KdV are indeed obtained from different
reductions of $N=4$ KdV.\par
The construction based on the abelian $u(1)^{\otimes 4}$ algebra could lead to
global $N=4$ hierarchies realized through strictly chiral and antichiral
superfields, but it can be easily checked that they are definitely not
polynomial generalization of
$N=2$ NLS and are not $N=4$ superconformal.

\section{Conclusions}

We have pointed out that supersymmetrical integrable hierarchies can be very
naturally investigated (and hopefully classified) taking as a starting point
the (super-)Lie algebras and their supersymmetric affinizations. Our approach
is very much complementary with the point of view advocated by many authors
in literature (like e.g. Z. Popowicz who is also author of a package for
computing Lax operators by using reduce).
They rather use the converse attitude of actually producing
integrable equations in terms of some consistent Lax operators, especially
of scalar type. This approach has the advantage of furnishing indeed
integrable systems, but leave aside questions concerning the algebraic
interpretation of these results. The approach based on Lie algebras has just
opposite merits and drawbacks. It furnishes from the very beginning the
algebraic
setting for defining dynamical systems and provides guidelines how to obtain
them, while the burden is on proving the
existence of a tower of hamiltonians in
involution.\par
This situation is very specific to the supersymmetric case since,
in
contrast with bosonic hierarchies, we do not dispose of a hamiltonian reduction
procedure which automatically leads to Lax operators. The examples where this
is
indeed the case, corresponding to the Inami-Kanno hierarchies, are of interest
but they belong to a restricted class. Other interesting super-integrable
systems like the $N=4$ KdV equation previously discussed are left out of this
scheme.\par
The approach based on (super-)Lie algebras
is a very convenient one in the investigation of supersymmetric extended
hierarchies. As discussed in this paper, one has to look for algebras admitting
extra structures, complex,
quaternionic and so far. Some partial results obtained
in collaboratin with Ivanov and Krivonos show
indeed that $sl(3)$, the next simplest quaternionic algebra, admits a global
$N=4$ structure which suggests the realization of at least global $N=4$
hierarchies.\par
In conclusion it deserves being mentioned that investigating supersymmetric
integrable hierarchies looks promising due to the presence of open problems.

\section{Acknowledgments}

This work has been done under JSPS contract. The author expresses its
gratitude to the members of the Phys. Dept. of Shizuoka University for their
kind hospitality.
%
%

\end{document}